\documentclass[12pt]{article}
\usepackage{amsmath, amsfonts,euscript,bbm}
\usepackage{epsfig, cite}
\usepackage{color}
\usepackage{ifthen}
\usepackage{graphicx}
\usepackage{setspace} 
\newboolean{Notes}\setboolean{Notes}{true}
\newcommand{\notes}[1]{\ifthenelse{\boolean{Notes}}{\textcolor{black}{#1}}{}}

\def \ha{{\textstyle{1\over 2}}}
\def \slope{{\textstyle{\alpha' \over 4}}}

\newcommand{\skyp}[1]{}

\def\lsim{\stackrel{<}{{_\sim}}}

\begin{document}

\bigskip
\hskip 4in\vbox{\baselineskip12pt \hbox{FERMILAB-PUB-06-288-A} }
\bigskip\bigskip\bigskip\bigskip\bigskip\bigskip\bigskip\bigskip

\centerline{\Large Cosmic Superstring Scattering in Backgrounds}
\bigskip
\bigskip
\bigskip
\centerline{\bf Mark G. Jackson}
\medskip
\centerline{Particle Astrophysics Center}
\centerline{Fermi National Accelerator Laboratory}
\centerline{Batavia, IL 60510}
\centerline{\it markj@fnal.gov}
\bigskip
\bigskip
\bigskip
\bigskip

\begin{abstract}
We generalize the calculation of cosmic superstring reconnection probability to non-trivial backgrounds.  This is done by modeling cosmic strings as wound tachyon modes in the 0B theory, and the spacetime effective action is then used to couple this to background fields.  Simple examples are given including trivial and warped compactifications. Generalization to $(p,q)$ strings is discussed.  
\end{abstract}

\newpage            
\baselineskip=18pt
\section{Introduction}
There has been a recent surge of interest in the possibility that superstrings may be observed as cosmic strings stretched across the sky \cite{Witten:1985fp} \cite{Polchinski:2004ia} \cite{Copeland:2003bj} \cite{Becker:2005pv}.  An important parameter in their observation is the intercommutation probability $P$, the likelihood that two cosmic strings approaching each other will reconnect to form a kinked shape.  This quantity determines  how many strings will be present in the present universe, and may also serve to differentiate what type of cosmic string it is \cite{Jackson:2004zg}.  Additionally, since superstrings are highly sensitive to any background present, they may also provide information about e.g. extra dimensions.  While order-of-magnitude estimates of how backgrounds would influence the intercommutation probability have been made, no rigorous technique has existed.  The purpose of the present article is to develop a simple method for calculating tree-level reconnection probability in slowly-varying backgrounds.  The driving motivation behind this study is that by examining the relationship between background and intercommutation probability, measurement of the latter could yield information about the former.  
\section{Interactions and Backgrounds}
After cosmic strings have formed through some symmetry-breaking mechanism, they do not sit idly.  Rather, their tension demands that they whip through space at nearly the speed of light, often colliding with themselves or other strings.  While gauge vortex cosmic strings intercommute with probability near unity \cite{matz}, for superstrings this is a scattering cross-section calculated by summing the relevant amplitudes: 
\[ P = \frac{1}{4 E_1 E_2 v} \sum_{\rm final} \left| \mathcal M _{\rm winding + winding \rightarrow kink} \right| ^2. \]
Rather than calculate this directly, for simple tree-level reconnection calculation it is more convenient to use a method introduced by Polchinski \cite{Polchinski:1988cn} whereby the optical theorem is used to express the answer in terms of the forward scattering amplitude:
\begin{equation}
\label{p}
P = \frac{1}{4 E_1 E_2 v} 2 {\rm Im} \left.  \mathcal M_{\rm winding + winding \rightarrow winding + winding} \right| _{t = 0} . 
\end{equation}
Thus we turn our attention to calculating the four-point winding string amplitude. 

It often happens that the energy scale needed to excite modes in certain dimensions is much greater than the energy scale $E$ under consideration.  This may be because some dimensions are physically very small,
\[ M_{\rm KK} \sim \frac{1}{R} \gg E, \]
or because there are background fields present which create a similar energy hierarchy \cite{DeWolfe:2002nn}, or even because the string carries an index in group space \cite{Hashimoto:2005hi}.  When one can make such a separation of scales, it is useful to `compactify' and average over these dimensions in the scattering process.  In the special case where the probability (\ref{p}) can be factorized into a noncompact part and a compact part, the latter acts effectively as a `volume factor' $V_{\rm \perp}$, diluting the probability:
\[ P=  \frac{1}{4 E_1 E_2 v V_{\rm \perp}} 2 {\rm Im} \ \mathcal M_{\rm 4D} . \]
Since the calculation of $\mathcal M$ in a general background is a nontrivial task, in \cite{Jackson:2004zg} an estimate of this effective volume was presented for strings in a six-dimensional confining background based on the scaling of string worldsheet interactions,
\begin{eqnarray*}
\frac{1}{V_{\rm \perp}} &\sim& \langle \delta ^6 (Y - Y') \rangle \\
&=& \frac{1}{(4 \pi)^3 \sqrt{\det \langle Y^i(0) Y^j(0) \rangle } } .
\end{eqnarray*}
A wound string in a confining background will acquire an effective worldsheet mass $m^2$ of order the background curvature, acting as an IR cutoff.  Assuming a 4D UV cutoff of order $\Lambda$, this implies a two-point function (no sum on $i$)
\begin{equation}
\label{twopoint}
 \langle Y^i(0) Y^i(0) \rangle \sim \frac{\alpha'}{2} \ln \left( 1 + \frac{\Lambda^2}{m^2} \right) .
 \end{equation}

Here we present a more rigorous method, using the spacetime effective action.  This allows one to easily include all slowly-varying background fields \cite{Callan:1986ja} \cite{deAlwis:1986rw} \cite{Tseytlin:1986tt}, and can in principle be carried out to any order in $\alpha'$ by including higher-derivative terms in the action.  The optical theorem is especially useful in this case because it allows one to compute the reconnection probability entirely within the framework of the effective action. 

 In the case where $\mathcal M$ is not factored the reconnection probability would depend non-trivially on the energy scale; investigation along these lines is in progress \cite{savmark}.  Both examples given here are factored. 

\section{0B Effective Action}
\subsection{Setup and Normalization}
We wish to study the interaction probability of cosmic superstrings which we model as wound fundamental strings.  These appear in the spectrum of all string theories except type I.  It would be easiest to perform the calculation in the type II theory, except that the lightest stable states have a polarization term in them making the scattering computation more tedious.  Since this would not affect the answer in the large-winding limit, we work with the technically simpler tachyon winding modes (this is of course a misnomer, as the field ceases to be tachyonic if the winding is more than a string length), which are usually removed out of the type II spectrum through the GSO projection.  To include these in the spectrum we perform a reversal of the usual GSO projection and thus consider the `type 0' string theories.  Tseytlin and Klebanov have studied the effective action of type 0B \cite{Klebanov:1998yy}, written here, keeping only the tachyonic terms relevant to our discussion:
\begin{equation}
\label{origeffaction}
S =  \frac{1}{4 \kappa^2} \int d^4 x d^6 y \sqrt{G} \left[ e^{-2 \Phi} \left( \ha G^{\mu \nu} \partial _\mu T \partial_\nu T  + \ha m^2_T T^2 \right) + W(T) \right]
\end{equation}
where $x$ represents the 4 noncompact dimensions, $y$ denotes the 6 compact dimensions, the type 0 tachyon mass is set by $m^2_T = -2/\alpha'$ (which is different from the bosonic tachyon mass), and $W(T)$ is the four-tachyon interaction term
\[ W(T) = \int \prod_{i=1} ^4 \frac{ d^D k_i}{ (2 \pi)^D} e^{i k_i x} W(k_1, k_2, k_3, k_4) T(k_1) T(k_2) T(k_3) T(k_4), \]
\[ W(k_1, k_2, k_3, k_4) = - 2 \pi \frac{\Gamma (-\slope s) \Gamma(- \slope t) \Gamma(-\slope u)}{\Gamma(1- \slope s) \Gamma(1+ \slope t) \Gamma(1+ \slope u)} . \]
The only part of $W$ that will be relevant is the large-energy forward scattering amplitude,
\[ W_{s \rightarrow \infty, t \rightarrow 0} \rightarrow - \frac{\pi \alpha'}{2} \frac{ s^2}{t} \left( \alpha' s/4 \right)^{\alpha' t/2} e^{-i \pi \alpha' t/4} . \]
The effective action also contains additional terms coupling the tachyon to both sets of RR fields (denoted $F_n$ and ${\bar F}_n$), an extra one appearing due to the GSO reversal.  Since the coupling requires both sets be nonzero, and one is not present in the IIB theory, we will not include these in our consideration.  Additionally, the $F_5 {\bar F_5} T$ coupling vanishes if $F_5$ is self-dual, which it is in the IIB theory.  This is all reasonable because fundamental strings should not couple to the RR fields.

We use this action to compute the invariant scattering matrix $\mathcal M$.  In the ten-dimensional case with only noncompact dimensions and no background, this is defined as
\[  i \mathcal M (2 \pi)^{10} \delta^{(10)} (\sum p)= \langle T_1, T_2 | e^{-iS_{\rm int}} | T_1,T_2 \rangle \]
where time-ordering is understood.  Upon compactification the $(2 \pi)^6 \delta^{(6)}(0)$ for the compact dimensions will be replaced with $V_{\rm eff}$.  The states $| T \rangle$ are eigenstates of the Hamiltonian, which in position represention means they satisfy the equation of motion
\[ \left( \nabla^\mu \nabla_\mu - 2 \nabla^\mu \Phi \nabla_\mu - m^2_T \right) \langle x ,y | T \rangle = 0 . \]
The normalization of $|T \rangle$ is fixed by the requirement that $\langle T_m | T_n \rangle = \delta_{mn}$, which after inserting the propagator from (\ref{origeffaction}) becomes
\begin{eqnarray*}
1&=& \frac{1}{4 \kappa^2} \int d^3 {\bf x} d^6 y \sqrt{G} e^{-2 \Phi} \langle T | x,y \rangle \frac{1}{2 E} \langle x,y | T \rangle \\
&=& \frac{1}{4 \kappa^2} \int d^3 {\bf x} d^6 y \sqrt{G} e^{-2 \Phi} \frac{ |T(x,y)|^2}{2E} .
\end{eqnarray*}
Since we are limiting ourselves to solutions which can be factored,
\[ T(x,y) = \phi(x) \psi(y), \]
the most convenient normalization is to demand
\begin{equation}
\label{psinorm}
1 = \int d^6 y \sqrt{G(y)} e^{-2 \Phi(y)} |\psi(y)|^2
\end{equation}
and so $\phi(x)$ will be normalized to its usual 4D value in the absence of any background:
\[ 1 = \frac{1}{4 \kappa^2} \int d^3 {\bf x} \frac{|\phi(x)|^2}{2E}. \]

\subsection{Winding modes}
\begin{figure}
\begin{center}
\includegraphics{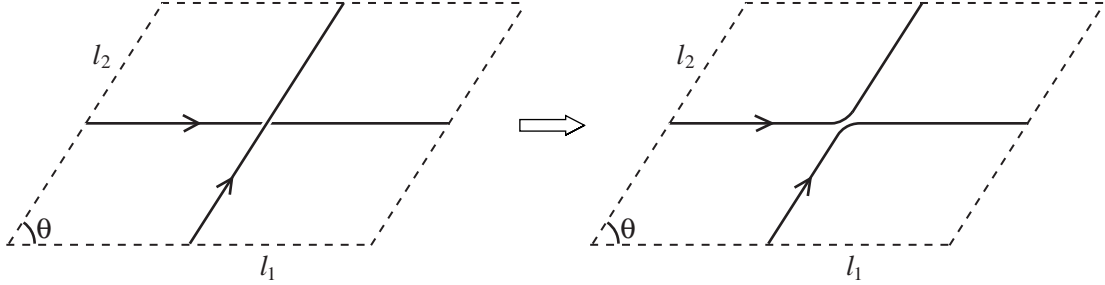}
\caption{Cosmic superstring scattering, whereby two straight wound strings reconnect into a single kinked closed string.} \ \\
\end{center}
\end{figure}
Now compactify $x_1, x_2$ on a torus of size $L$ and skew angle $\theta$ (see Figure 1), so that the tachyon field can be decomposed into winding modes:
\[ T = \sum_{n_1,n_2} T_{n_1,n_2} e^{i \left[ x_1 (n_1 + n_2 \cos \theta) + i x_2 n_2 \sin \theta \right] L / 2 \pi \alpha'} . \]
Since wound strings are charged under the Kalb-Ramond field $B_{\mu \nu}$ the derivative must become a gauge-covariant derivative, we can think of $B_{1\mu}$ and $B_{2 \mu}$ as two separate gauge fields $A^{(1)}_\mu$ and $A^{(2)}_\mu$ under which the transverse coordinates of the string are charged.  Couplings to the gauge fields must now be introduced,
\[ \partial_\mu \rightarrow D_\mu = \partial_\mu - i (n_1 + n_2 \cos \theta) B_{1 \mu} - i n_2 \sin \theta B_{2 \mu} . \]
The 0B effective action then becomes 8-dimensional with 2D compactification factor $L^2 \sin \theta$,
\begin{equation}
\label{compeffaction}
 S = \frac{L^2 \sin \theta}{4 \kappa^2} \sum_{n_1,n_2} \int dt dz d^6 y \sqrt{G} \left[ e^{-2 \Phi} \left( \ha |D _\mu T_{n_1,n_2} |^2 + \ha m^2_{n_1,n_2} |T_{n_1,n_2}|^2 \right) + W(T_{n_1,n_2}) \right]
 \end{equation}
where
\[ m^2_{n_1,n_2} = \left( \frac{L}{2 \pi \alpha'} \right)^2 \left[ (n_1 + n_2 \cos \theta)^2 + n_2 ^2 \sin ^2 \theta \right] + m_T^2 \]
is the mass of the winding string.  The free state solutions to the theory are now given by 
\[ D^\mu D_\mu T_{n_1,n_2} - 2 D^\mu \Phi D_\mu T_{n_1,n_2} = m^2_{n_1,n_2} T_{n_1,n_2}. \]

The quantity that we are interested in is the forward amplitude for strings wound once around each direction,
\[  \langle T_{1,0} ,  T_{0,1} |e^{-i S_{\rm int}} | T_{1,0}, T_{0,1} \rangle. \]
This amplitude can then be factored into a noncompact part, which has already been evaluated \cite{Jackson:2004zg} \cite{Polchinski:1988cn}, times an effective volume factor due to the compact wavefunction.  This allows us to define
\[  \mathcal M = \mathcal M_{4D} \int d^6 y \sqrt{G} e^{-2\Phi} \psi^*_{1,0} \psi^*_{0,1} \psi_{1,0} \psi_{0,1} \]
Substituting this into (\ref{p}) gives the reconnection probability
\[ P =  g_s^2 f(\theta,v) \frac{V_{\rm min} }{V_{\rm \perp}} \]
where
\[ V_{\rm min} = (4 \pi^2 \alpha')^3 , \hspace{0.5in} f(\theta,v) = \frac{(1 - \cos \theta\sqrt{1-v^2} )^2}{8 \sin \theta v \sqrt{1-v^2}}, \]
\[ \frac{1}{V_{\rm \perp}} =  \int d^6 y \sqrt{G} e^{-2\Phi} |\psi_{1,0}|^2 |\psi_{0,1}|^2 \]
and each $\psi(y)$ obeys the normalization condition (\ref{psinorm}).

\section{Examples}

\subsection{Simple Compactification}
Consider the situation where the compact six dimensions are a simple torus of volume $V$,
\[ \int d^6 y = (2 \pi R)^6 = V . \]
This must be chosen in an intermediate range so that we can both ignore the KK modes ($R \lsim \sqrt{\alpha'}$) and also trust the effective action ($(4 \pi^2 \alpha')^3/V \ll 1$).  The propagator volume integral will normalize the compact wavefunction to $\int d^6 y |\psi(y)|^2 = 1$, so assuming that there is no momentum in these dimensions and $\psi$ is constant, the solution for each is simply
\[ \psi(y) = \frac{1}{\sqrt{V}} . \]
Performing the volume integral in the compact amplitude then gives the expected result
\[ \frac{1}{V_{\rm \perp}} = \int d^6 y |\psi(y)|^4 = \frac{1}{V} \]
confirming that the intercommutation probability is suppressed by a factor of $1/V$.
\subsection{Warped Compactification}
Now consider a less trivial example, whereby the volume of the ``compact" dimensions is large but the background localizes the string wavefunction to produce an effective volume (see Figure 2).  A well-studied example of warped compactification is the Klebanov-Strassler geometry \cite{Klebanov:2000hb} which has been used in the KKLMMT model of brane inflation \cite{Kachru:2003sx}.  In this model the local geometry is $\mathbb R ^3 \times S^3$ with the warp factor a function only of $r^2 = \sum_{i=1} ^3 r_i^2$:
\[ ds^2 = e^{A(r)} \eta_{\mu \nu} dx^\mu dx^\nu + e^{-A(r)} \left( dr_1^2+ dr_2^2 + dr_3^2 + R_3^2 d\Omega^2_3 \right) . \]
\begin{figure}
\begin{center}
\includegraphics{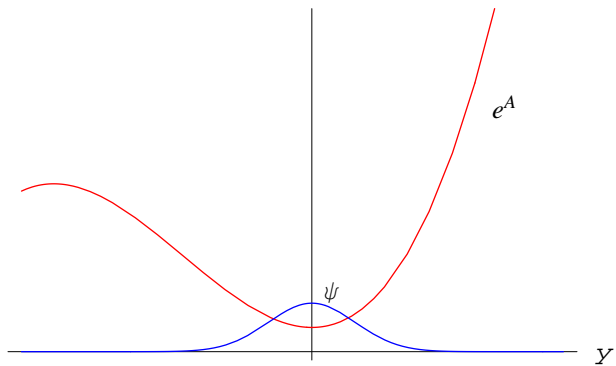}
\caption{Any local minima in the warp factor $e^A$ (red) can be approximated as a quadratic potential, which will have a gaussian ground state (blue).} \ \\
\end{center}
\end{figure}
The dilaton variation and Kalb-Ramond field are set to zero in this background.  The wavefunction is then given by (we will suppress $n_1,n_2$ indices)
\[ \nabla^M \nabla_M T(x,y) = m^2 T(x,y) . \]
The interaction with the background comes from the metric connection terms.  Approximating the warp factor near the bottom of the throat as a quadratic potential, 
\[ e^A \approx a_0 + a_1 r^2, \]
these combine to produce the equation of motion ($\mu$ denotes the four Minkowski dimensions, $i$ denotes $r^i$ on the $\mathbb R^3$, $a$ denotes the angular coordinates on the $S^3$)
\[ \left[ e^{-A} \partial^2_\mu + e^A \left( \partial_i^2 - 5a_1 x^i \partial_i   + \frac{1}{R^2_3} \partial_a^2\right)  \right] T(x,y) = m^2 T(x,y) . \]
We now further approximate $e^A \approx a_0 \ll 1$, allowing us to factor $T$ into parts which separately obey the equations
\begin{eqnarray*}
\partial_\mu ^2 \phi(x) &=& \left( a_0 m^2 + \frac{\epsilon}{a_0} \right) \phi(x) , \\
- \left[ \partial_i^2 - 5a_1 x^i \partial_i   + \frac{1}{R^2_3} \partial_a^2 \right] \psi(y) &=& \frac{\epsilon}{a_0} \psi(y).
\end{eqnarray*}
The first is the usual 4D wave equation but with mass rescaled due to the warping, plus the KK mass term due to compactification.  The second is a Schrodinger equation for the compact space wavefunction.  The Laplacian on the $S^3$ will have eigenvalue $-l(l+2)$; we will assume that the string remains in the ground state $l=0$, and so has the 3D spherical harmonic $Y_{1,0,0} (y^a) = 1/\sqrt{2 \pi^2}$.  The operator acting on the $\mathbb R^3$ has solutions which are isomorphic to the simple harmonic oscillator which we take to be in the gaussian ground state.  The complete normalized compact wavefunction is then
\[ \psi(y) = \sqrt{ \frac{a_0}{ 2 \pi^2 R_3^3} } \left( \frac{5a_1}{2 \pi} \right)^{3/4} e^{-5a_1 r^2/2}. \]
This yields an effective volume
\[ \frac{1}{V_{\rm \perp}} = \int d^6 y \sqrt{G} |\psi|^4 =  \frac{a_0}{ 2 \pi^2 R_3^3}  \left( \frac{5a_1}{2 \pi} \right)^{3/2} . \]
Thus $P$ increases as $a_1$ gets larger and the wavefunction is more confined.  This agrees with the previous qualitative estimate (\ref{twopoint}) in the long-wavelength limit of $(\Lambda/m)^2=5a_1 \alpha'  \ll 1$.   The warping parameters for the KKLMMT inflation model are then determined by the scale of inflation and the flux integers,
\[ a_0 \sim 10^{-2}, \hspace{0.5in} a_1 \sim \frac{a_0}{2 g_s M \alpha'}, \hspace{0.5in} R_3 \sim \sqrt{g_s M \alpha' }. \]
Combining these factors we have an effective volume
\[ \frac{1}{V_{\rm \perp}} \sim \frac{ 5^{3/2} \ 10^{-5} }{16 \pi^{7/2} (g_s M \alpha')^3} . \]

\section{Discussion and Conclusion}
In this article we have generalized the computation of basic cosmic superstring reconnection probabilities for a general background.  We have considered only very simple examples but it would be instructive to study more sophisticated backgrounds and to higher order in $\alpha'$.  It would be especially interesting to consider nonfactored amplitudes, as this would allow a sensitive probe of the extra dimensions at energies inaccessible to accelerators.

Ideally we would be able to use the same technique to calculate the effect of the background for all types of cosmic superstrings, fundamental as well as $(p,q)$ \cite{Schwarz:1995dk}; as it stands, the methods are completely different \cite{Jackson:2004zg} \cite{Hanany:2005bc}.  Though we have used the effective action of the perturbative 0B string theory, it is known that IIB can accomodate the nonperturbative $(p,q)$ strings, so we conjecture that the 0B action should be capable of this in our approach.  The main difference would seem to be that the tachyon would couple to both the NSNS and RR two-forms, though it is unclear precisely how this would work since there are now two sets of RR fields.  In this case the wavefunction for the $(p,q)$ string is to leading order in $g_s$ simply $\psi(y) = \delta(y-y_0)$, where $y_0$ will depend on the charges $p$ and $q$ coupled to the two-form background fields.

Finally the effect of the background may also prove important in string and brane gas studies \cite{Easther:2002mi} \cite{Easther:2004sd}, where the decompactification rate of the universe depends on winding mode interaction rates.
\section{Acknowledgements}
I would like to thank J. Lykken, J. Polchinski and S. Sethi for useful discussions, and I am supported by NASA grant NAG 5-10842.


\end{document}